# SPACE WEATHER RESEARCH AND FORECAST IN USA

Pevtsov, A.A.

*National Solar Observatory, Boulder, Colorado 80303, USA*

*In the United States, scientific research in space weather is funded by several Government Agencies including the National Science Foundation (NSF) and the National Aeronautics and Space Agency (NASA). For commercial purposes, space weather forecast is made by the Space Weather Prediction Center (SWPC) of the National Oceanic and Atmospheric Administration (NOAA). Observations come from the network of groundbased observatories funded via various sources, as well as from the instruments on spacecraft. Numerical models used in forecast are developed in the framework of individual research projects. Later, the most promising models are selected for additional testing at SWPC. In order to increase the application of models in research and education, NASA in collaboration with other agencies created Community Coordinated Modeling Center (CCMC). In mid-1990, US scientific community presented compelling evidence for developing the National Program on Space Weather, and in 1995, such program has been formally created. In 2015, the National Council on Science and Technology issued two documents: the National Space Weather Strategy [1] and the Action Plan [2]. In the near future, these two documents will define the development of Space Weather research and forecasting activity in USA. Both documents emphasize the need for close international collaboration in area of space weather.*

INTRODUCTION

At the time of writing this review, the space weather is taking the society by storm: in US, at the end of 2015, the National Science and Technology Council (NSTC) issued two important documents [1, 2] that will define the future of space weather development, in U.K. "The National risk register for civil emergencies" [3] identified the space weather as one of major risk factors. Space weather prediction centers have been established in several countries. The effects of solar activity on Earth (or solar-terrestrial connections) were discovered long time ago, shortly after the discovery of cycle variations in solar (sunspot) activity. The very first solar flare, observed in 1859 by Richard Carrington (and independently, by Richard Hodgson) also caused the very first space weather event: wide spread aurora borealis, strong variations in geomagnetic field and significant electric currents induced in telegraph wires across several continents. The earliest forecasts of solar activity for societal benefit begun in $1930^{th}$-$1940^{th}$ mainly for forecasting the conditions for radio communications at high latitudes (near Polar regions). After the World War II, forecast of solar activity was used by military and civilian customers (e.g., to anticipate the negative effects on radar detection systems and early navigation systems: LORAN, OMEGA (USA) and CHAIKA, ALPHA (USSR), and later, in support of human space exploration). A true need for having a reliable forecast of solar and geomagnetic activity was highlighted by a military incident, which nearly started a nuclear war, when the effects of a major solar flare resulted in a catastrophic failure of the US early warning system [4]. After that incident, the operational forecast of solar activity was performed both in US and USSR. Figure 1 show a timeline, when the forecast for specific area of activity was first started.

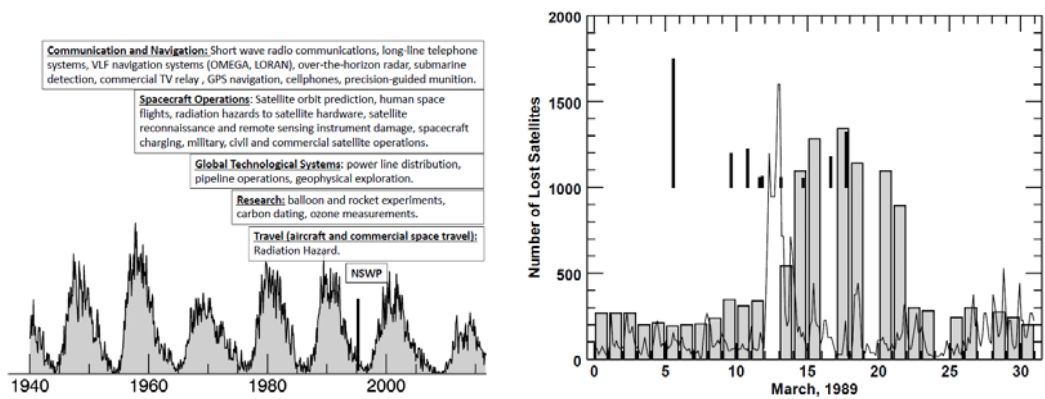

Figure 1. (left): List of activities currently relying on space weather forecast. Starting position of each box corresponds to approximate year (shown at the bottom) when the forecast started for that group. For a reference, solid line with gray halftone shows international sunspot number. Thick vertical line in 1995 indicates the year when the US National Space Weather Program (NSWP) was formally established. (right): Number of orbital objects that are not on the expected orbits (shaded bars) during March 1989. Thick vertical lines mark time and importance of solar X-ray class flares (tallest line corresponds to X-class flare), thin black line shows the geomagnetic Ap-index. For color version see doi: 10.7910/DVN/L3HCC6.

Catastrophic events related to major space weather disturbances are well-documented. Those include, for example, a premature loss of Skylab due to higher-then-expected atmospheric drag (which was the result of enhanced level of solar activity), loss of several satellites due to electric discharge (e.g., Canadian Anik E1 and E2 communication satellites); massive blackouts due to electric power grid collapse (e.g., PowerQuebec system). More important, however, is that there are significant impacts from smaller "non-catastrophic" events. Figure 1 (right) show example of a significant increase in number of orbital objects that change their orbits significantly (so called, lost satellites) due to change in the atmospheric drag associated with major X-ray flare and geomagnetic storm. Some of these "lost satellites" are orbital debris, which may threaten functioning satellites or even the International Space Station. To plan for the avoidance maneuvers require re-calculating the orbital parameters of these "lost satellites". There could also be significant cost in operational loses for services that rely on global navigation systems (GPS, GLONASS) or conduct operations in Earth's Polar Regions (e.g., inter-continental polar flights).

FORECASTING CENTERS

In US, there are two centers for operational space weather forecast: Space Weather Prediction Center (SWPC, Boulder, Colorado) and the prediction center run by the US Airforce 557$^{th}$ Weather Wing (Omaha, Nebraska). Two centers work in close coordination. SWPC is the Nation's official source of space weather watches, warnings and alerts. In evaluating the space weather environment, SWPC uses observational data from groundbased and spaceborne instruments including NOAA satellites (the Geostationary Operational Environmental Satellite, GOES, the Polar-orbiting Operational Environmental Satellite, POES, Deep Space Climate Observatory, DSCOVR); NASA research satellites (Advanced Composition Explorer, ACE; Solar Dynamics Observatory, SDO; Solar Terrestrial Relations Observatory, STEREO, and NASA/ESA's Solar and Heliospheric Observatory, SOHO). Groundbased observations come from 6-instrument GONG network (NSO/NSF/USAF, full disk magnetograms and H$_\alpha$ images), the US Airforce's Solar Observing Optical Network (SOON, sunspot and flare information) and the Radio Solar Telescope Network (RSTN, radio bursts). For GOES, DSCOVR, and POES satellites, NOAA operates its own ground-support station, with additional support from the Air Force Satellite Control Network (AFSCN); NASA's Wallops Command and Data Acquisition Station (WCDAS) in Virginia; National Institute of Information and Communications Technology (NICT) in Tokyo, Japan; Korean Space Weather Center (KSWC) in Jeju, Korea; German Aerospace Center (DLR) in Neustrelitz, Germany. Data from NASA and ESA satellites are downloaded via NASA/ESA run groundbased stations. SWPC is part of the International Space Environment Service (ISES, [5]), with regional centers located in Australia (Sydney), Austria

(Treffen), Belgium (Brussels), Brazil (São José dos Campos), Canada (Ottawa), China (Beijing) , Czech Republic (Prague), India (New Delhi), Indonesia (Jakarta), Japan (Tokyo), Mexico (Morelia), Poland (Warsaw), Russia (Moscow), South Africa (Hermanus), South Korea (Jeju), Sweden (Lund), UK (Exeter), and USA (Boulder). In addition, the CLS warning center in France (Toulouse) is affiliated with ISES through the Regional Warning Center in Brussels, Belgium; three additional ISES-associated centers are located in China (Beijing). In ISES structure, SWPC "plays a special role as "World Warning Agency", acting as a hub for data exchange and forecasts."[5]

SWPC OPERATIONS AND DATA PRODUCTS

   As part of its daily operations, SWPC provides such information to its customers and general public as forecasts; reports; models; observations; summaries; and alerts, watches and warnings. Category "forecasts" includes: 27-Day Outlook of 10.7 cm Radio Flux and Geomagnetic Indices; 3-Day Forecast; 3-Day Geomagnetic Forecast; Predicted Sunspot Numbers and Radio Flux; Report and Forecast of Solar and Geophysical Activity; Solar Cycle Progression; Space Weather Advisory Outlook; USAF 45-Day Ap and F10.7cm Flux Forecast; Weekly Highlights and 27-Day Forecast. Category "reports" includes: Forecast Verification; Geoalert - Alerts, Analysis and Forecast Codes; Geophysical Alert; Solar and Geophysical Event Reports; USAF Magnetometer Analysis Report. Category "models" includes: Aurora - 30 Minute Forecast; D Region Absorption Predictions (D-RAP); North American Total Electron Content (US and non-US regions); Relativistic Electron Forecast Model; STORM Time Empirical Ionospheric Correction; WSA-Enlil Solar Wind Prediction; USAF Wing model Kp predicted activity index. Category "observations" contains: ACE Real-Time Solar Wind; Boulder Magnetometer; GOES Data (Electron Flux, Magnetometer, Proton Flux, Solar X-ray Imager, X-ray Flux); LASCO Coronagraph Images; Planetary K-index; Real Time Solar Wind, Satellite Environment Data; Solar Synoptic Maps (created and used by SWPC forecasters); Station K and A Indices. "Summaries" provide detailed descriptions of overall space weather activity, and "alerts, watches and warnings" provide brief summaries of space weather activity, impact on technological systems, and expected scale of this impact.
   To assist customers in their decision making process, SWPC developed a simple classification matrix that represents potential scale of the event. For example, solar radiation storms (forecast is designated by letter "S") could be categorized between minor (S1) and extreme (S5). Other letter categories are radio blackouts (R) and geomagnetic storms (G). For each of these letter categories, SWPC provides a brief summary that explains the scale of the event, its potential impact, the physical measure (in units of X-ray flux, Kp index or particle energy range) and the frequency of events.  In addition, SWPC provides several experimental data products, which currently include: Aurora - 3 Day Forecast; CTIPe Total Electron Content Forecast; Geospace Equatorial and Meridional Magnetospheric Views; Geospace Global Geomagnetic Activity Plot; Geospace Ground Magnetic Perturbation Maps. Examples of some SWPC data products are shown in Figure 2. The forecast information is provided at the SWPC web site [6], and selected targeted information is also distributed directly to the subscribers, which include satellite operators, shipping, banking, airline industry, communication companies, oil drilling, electric utilities, precision agriculture, surveying groups, US Department of transportation, FEMA, FAA, manned space flight, and academia. The subscription to space weather forecast services was opened in January 2005, and as of end of 2015, the total number of subscribes was 47,131.

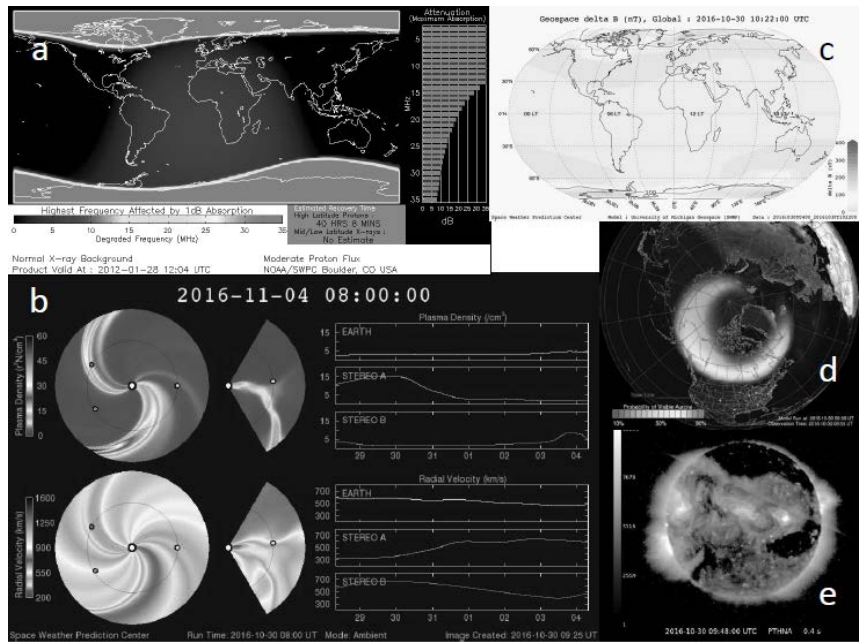

Figure 2. Example of SWPC data products: (a) D Region Absorption Predictions (D-RAP); (b) WSA-Enlil Solar Wind Prediction; (c) Geospace Ground Magnetic Perturbation Map, (d) Aurora - 30 Minute Forecast, (e) coronal image from GOES Solar X-ray Imager. Color version of this figure can be found at **doi:10.7910/DVN/L3HCC6**.

RESEARCH IN SPACE WEATHER

Computer models and empirical relationships for space weather forecast are developed in the framework of individual research projects supported via grant programs mostly by NSF and NASA. Normally, these research projects aim at scientific understanding of physical processes not the model development. The models and empirical relations used in space weather are usually a bi-product of these projects. As the new relationships discovered and verified, and the numerical models mature, they can be selected for additional testing at SWPC. In 1998, to stimulate a broader use of space science and space weather models in scientific research, and to promote development of next-generation of models, NSF and NASA created a Community Coordinated Modeling Center (CCMC) at NASA's Goddard Space Flight Center. Current CCMC activity helps in broadening the user base for numerical modeling of solar and heliospheric phenomena, promotes developing interfaces between different models, employs scientific models for teaching purposes, and provides opportunity collaboration between modelers and additional testing of models.

FUTURE DEVELOPMENTS

In the near future, [1,2] will define the development of Space Weather research and forecasting activity in US. The documents emphasize that the space weather is a global challenge, which requires strong international collaboration. Such collaboration can include increasing "engagement with the international community on observation infrastructure, data sharing, numerical modeling, and scientific research", strengthening "international coordination and cooperation on space-weather products and services", and promoting "a collaborative international approach to preparedness for extreme space-weather events." These opportunities for establishing broader collaboration between international groups involved in space weather research and forecast need to be fully exploited. For example, there are discussions in the research community about the benefits of launching a spacecraft at Lagrangian L5 point (e.g., U.K. Carrington project) and developing a new SPRING network of solar groundbased telescopes to replace ageing GONG network. Russian scientific community may consider participating in some of these initiatives, and maybe even promote having its own spacecraft at Lagrangian L4 as a contribution to L5 mission.